\documentclass[conference]{IEEEtran}
\usepackage{cite}

\newtheorem{definition}{Definition}
\usepackage{adjustbox}
\usepackage{algorithm}
\usepackage{algorithmicx,algpseudocode}
\usepackage{graphicx}
\usepackage{float}
\usepackage{amsmath}
\usepackage{epstopdf}

\usepackage{setspace}

\algnewcommand{\LineComment}[1]{\State \(\triangleright\) #1}

\begin{document}
\title{Towards Unified Combinatorial Interaction Testing}

\author{
Hanefi Mercan and Cemal Yilmaz\\
\textit{Faculty of Engineering and Natural Sciences, Sabanci University, Istanbul, Turkey}\\
\textit{\{hanefimercan,cyilmaz\}@sabanciuniv.edu}
}

\date{Today}
\maketitle

\begin{abstract}
We believe that we can exploit the benefits of combinatorial interaction testing (CIT) on many ``non-traditional'' combinatorial spaces using many ``non-traditional'' coverage criteria. However, this requires truly flexible CIT approaches. To this end, we introduce {\em Unified Combinatorial Interaction Testing} (U-CIT), which enables practitioners to define their own combinatorial spaces and coverage criteria for testing, and present a unified construction approach to compute specific instances of U-CIT objects. We, furthermore, argue that most (if not all) existing CIT objects are a special case of U-CIT and demonstrate the flexibility of U-CIT on a simple, yet realistic scenario.
\end{abstract}

\begin{keywords}
combinatorial interaction testing; covering arrays; satisfiability problem
\end{keywords}

\section{Introduction} 
\label{intro}

Software systems frequently embody a wide spectrum of system variabilities that require testing, such as software and hardware configuration options, user inputs, thread interleavings, sequence of events/operations,  or software product families. However, exhaustively testing all possible variations in a timely manner (if not impossible at all) is generally far beyond the available resources~\cite{yilmaz2013moving}.


For this reason, the testing of industrial systems almost
always involve sampling enormous variability spaces and testing
representative instances of a system's behavior. In practice, this sampling is commonly performed with techniques collectively referred to as combinatorial interaction testing, (or CIT) \cite{yilmaz2013moving,nie2011survey}. CIT typically models a system under test (SUT) as a set of factors (choice points, parameters, or configuration options), each of which takes its values from a particular domain. Based on this model, CIT then generates a sample, meeting a specified \emph{coverage criterion}. That is, the sample contains some specified combinations of the factors and their values. For instance, a $t$-way covering array, which is a well-known and frequently-used CIT object, requires that each valid combination of factor values for every combination of $t$ factors, appears at least once in the sample~\cite{cohen1997aetg}.

The basic justification for using CIT is that they can (under certain assumptions) effectively and efficiently exercise all system behaviors caused by the interactions of $t$ or fewer factors. The effectiveness of CIT stems from the coverage it provides; e.g., all required $t$-way combinations of factor values are guaranteed to be covered at least once. The efficiency, on the other hand, stems from the fact that a test case can cover more than one required combinations. Therefore, carefully generating the test cases, such that a full coverage under the given criterion is obtained using a minimum number of test cases, e.g., can decrease the cost of testing.

The results of many empirical studies suggest that majority of factor-related failures in practice are caused by the interactions of only a small number factors. That is, $t$ is small in practice, typically $2 \leq t \leq 6$ with $t$$=$$2$ (i.e., pairwise testing) being the most common case~\cite{cohen1996combinatorial,cohen1997aetg,czerwonka2006pairwise,dalal1998model}. For a fixed $t$, as the variability space grows (as the number of factors increases, for example), the size of CIT objects represents an increasingly smaller proportion of the whole space. Thus, very large spaces can be efficiently covered. Consequently, CIT has been successfully used in many domains,  including systematic testing of protocols~\cite{burroughs1994improved,williams1996practical}, input parameters~\cite{schroeder2002generating}, configurations~\cite{kuhn2008practical,yilmaz2006covering}, software product lines~\cite{johansen2012algorithm}, multi-threaded applications~\cite{lei2007combinatorial}, and graphical user interfaces~\cite{yuan2011gui}. 
 
All these have so far been achieved by having researchers develop  specific models for defining variability spaces together with specific coverage criteria for testing, both of which are then used by practitioners. We believe that we can further exploit the benefits of CIT on many ``non-traditional'' variability spaces using many ``non-traditional'' coverage criteria~\cite{yilmaz2013moving}. However, this requires truly flexible CIT approaches. To this end, we conjecture that the flexibility, thus the applicability of CIT in practice, would greatly be improved, if there were better tools that allowed practitioners to define their own application-specific variability spaces as well as their own application-specific coverage criteria. That is, rather than we, as researchers, invent new CIT objects for testing and ask practitioners to use them (thus telling them what to test), we would like to enable practitioners to define their own space for testing as well as their own coverage criterion (thus enabling them to invent their own application-specific CIT objects). Our goal as researchers would then be to develop powerful tools to efficiently and effectively sample the given space to obtain full coverage under the given criterion. Although such generic tools may not be as efficient as their specialized counterparts, they certainly can provide the flexibility needed in practice.

In this work we first informally introduce {\em Unified Combinatorial Interaction Testing} (U-CIT), which enables practitioners to define their own variability spaces and coverage criteria for testing, and present a unified construction approach to compute specific instances of U-CIT objects. We then argue that most (if not all) existing CIT objects are a special case of U-CIT by informally specifying two well-known CIT objects, namely traditional covering arrays and sequence covering arrays~\cite{kuhn2012combinatorial}, as U-CIT objects. Finally, we demonstrate the flexibility of U-CIT on a simple, yet realistic scenario, which existing CIT objects suffer to address.

\section{Unified Combinatorial Interaction Testing (U-CIT)}
\label{ucit}

At a high level, U-CIT takes as input a specification that implicitly defines a space of all valid test cases and a coverage criterion that implicitly defines all entities that need to be covered by testing. The output is an object, e.g., a set of valid test cases, which achieves a full coverage under the given criterion. Although it is possible to define additional constraints on the emergent properties of the resulting objects, such as the objects must achieve a full coverage with the ``minimum'' possible testing cost~\cite{demiroz2012cost}, we, for this work, assume one such  constraint which aims to minimize the number of test cases required for full coverage.

What makes a U-CIT approach a unified approach is that entities to be covered, test cases, and the space from which the test cases will be sampled, are all expressed as constraints. Consequently, the problem of computing a U-CIT object turns into one big, interesting constraint solving problem. Note that we use the term ``constraint'' in the general sense; any restriction, independent of the logic in which it is specified, is considered to be a constraint. In other words, no matter whether the constraints are specified using Boolean logic, first-order logic, temporal logic, etc., the proposed approach will work as long as an appropriate constraint solver is provided.
 
In particular, a coverage criterion implicitly defines a set of constraints to be satisfied (not necessarily all together, but in groups), where each constraint represents an entity to be covered. A U-CIT object is then computed by finding a ``minimum'' number of subsets of these constraints, such that within a subset all constraints are satisfiable together and that the union of all subsets is the same as the original set of constrains inferred from the coverage criterion. In effect, a solution to a subset of constraints represents a valid test case, i.e., a collection of entities that can be tested together. Therefore, a set of test cases generated for all the subsets on a one test case per subset basis, represents a U-CIT object achieving full coverage under the given coverage criterion. Note that the specification of the variability space further constraints the entities to be covered as well as the test cases to be sampled. More formally:

\begin{definition}
A {\em U-CIT requirement} is an entity expressed as a constraint, which needs to be covered at least one U-CIT test case. 
\end{definition}

\begin{definition}  
A {\em U-CIT test case} is a collection of U-CIT test requirements that can be tested together, i.e., a set of constraints that can be satisfied together.  
\end{definition}

Not all possible combinations of requirements may be valid in practice.

\begin{definition}
A {\em U-CIT space model} is a system of constraints that implicitly define the space of all valid U-CIT requirements as well as all valid U-CIT test cases. 
\end{definition}
 



\begin{definition}
A {\em U-CIT coverage criterion} is a criterion that implicitly defines a set of U-CIT requirements to be covered.
\end{definition}

\begin{definition}
A {\em U-CIT object} is a collection of U-CIT test cases, which achieves a full coverage under a given U-CIT coverage criterion, i.e., a collection of test cases, in which for every requirement specified by the coverage criterion, there is at least one test case, in which the respective constraint is satisfied.
\end{definition}

\section{Specifying Existing CIT Objects as U-CIT Objects}
\label{example}

In this section we, as an example, informally specify two well-known CIT objects, namely traditional covering arrays and sequence covering arrays, as U-CIT objects using the definitions given in Section~\ref{ucit}.
  
\subsection{Traditional Covering Arrays}
\label{ca}

Let a $t$-tuple be a set of factor-value pairs for a combination of $t$ distinct factors. Then, given a coverage strength $t$ and a configuration space model comprised of a set of factors, each of which takes its values from a discrete domain, together with a system-wide constraint (if any), which invalidates certain combinations of factor values, a $t$-way covering array is a set of valid $n$-tuples (i.e., a set of test cases), in which every valid $t$-tuple appears at least once, where $n$ is the number of factors in the configuration space model.

One way of specifying $t$-way traditional covering arrays as U-CIT objects is:

\begin{itemize}

\item {\em A U-CIT requirement:} A constraint that represents a valid $t$-tuple.

\item {\em A U-CIT test case:} A valid $n$-tuple.

\item {\em U-CIT space model}: A constraint system specifying that 1) every factor takes its value from a particular discrete domain, 2) a valid U-CIT requirement is a $t$-tuple that does not violate the system-wide constraint, and 3) a valid U-CIT test case is an $n$-tuple that does not violate the system-wide constraint.

\item {\em U-CIT coverage criterion:} All valid $t$-tuples must be covered at least once. Determining all valid $t$-tuples can trivially be performed (Section~\ref{Approach}).

\item {\em U-CIT object:} A traditional $t$-way covering array.

\end{itemize}

Note that all of the constraints discussed above can trivially be specified in Boolean logic or first-order logic and then solved using a general-purpose constraint solver.

\subsection{Sequence Covering Arrays}
\label{sca}

With traditional covering arrays, order of factor values in a given test case is assumed to have no effect on the fault revealing ability of the test case. This assumption, however, may not always hold true in practice. For example, in event-driven systems, such as found in graphical user interfaces and device drivers, the way an event is processed often depends on the sequence of preceding events. Therefore, different orderings of the same set of events can reveal different failures.

To address these issues, sequence covering arrays are built to cover orderings of events. For a given set of $n$ events together with a system-wide constraint (if any), which invalidates certain event orderings, a $t$-way sequence covering array is a set of sequence of events, in which every possible event sequence of length $t$ appears at least once~\cite{kuhn2012combinatorial}, while the events in the sequence can be interleaved with other events. Different variations of these objects exist. In one variation, for example, each event sequence computed as a test case, must be a permutation of all events, i.e., each event must appear exactly once in the sequence. In another variation, not all events are required to appear in a test case and/or the same event can appear multiple times.

One way of specifying $t$-way sequence covering arrays as U-CIT objects is:

\begin{itemize}

\item {\em A U-CIT requirement:} A constraint that represents a valid $t$-length sequence of events.

\item {\em A U-CIT test case:} An $n$-length or a variable-length sequence of events. 

\item {\em U-CIT space model}: A constraint system specifying that 1) a valid U-CIT requirement is a $t$-length sequence of events that does not violate the system-wide constraint and 2) a valid U-CIT test case is either an $n$-length or a variable-length sequence of events (depending on the variation to be used),  which does not violate the system-wide constraint.
 
\item {\em U-CIT coverage criterion:} All valid $t$-length event sequences must be covered at least once. Determining all valid $t$-length event sequences can trivially be performed (Section~\ref{Approach}).

\item {\em U-CIT object:} A $t$-way sequence covering array.

\end{itemize}

These constraints can be expressed in Boolean and/or first-order logic and solved using a general-purpose constraint solver.

\begin{algorithm}[t]
\small
{\bf Input:} A U-CIT space model $M$ \\
{\bf Input:} A set of U-CIT requirements $R$ to be covered\\
{\bf Output:} A U-CIT object $S$ \\ 
\begin{algorithmic}[1]
\State 
\State $S \leftarrow \{\}$  
\For { {\bf each} $r \in R$}
	\State $accommodated \leftarrow false$
	\For {{\bf each} $R' \in S$}
		\If{$ isSatisfiable(r \wedge M \wedge \bigwedge_{r' \in R'}r')$} 
			\State $R' \leftarrow R' \cup r$
			\State $accommodated \leftarrow true$
			\State \textbf{break}
		\EndIf			\EndFor
	\If{ {\bf not} $accommodated$}  
	    \State $S \leftarrow S \cup \{r\}$     
	\EndIf		
\EndFor
\State \textbf{return} $S$ 
\end{algorithmic}
\caption{An algorithm for computing U-CIT objects}
\label{algo:approach1}
\end{algorithm}
\section{An Approach for Computing U-CIT Objects} 
\label{Approach}

In this section we present a proof-of-concept approach to compute U-CIT objects. 

Given a U-CIT space model $M$, which is indeed a constraint system and a coverage criterion $C$, we first determine all valid U-CIT requirements. To this end, we enumerate all the entities to be covered, convert each entity to a constraint $r$, and then determine whether $r \wedge M$ is satisfiable. If it is, then $r$ is a valid requirement. Otherwise, $r$ is invalid. 

Once the set of valid requirements $R$ are determined, we use a greedy algorithm (Algorithm 1) to compute a ``minimum'' number of satisfiable subsets of $R$, such that each and every valid requirement appears in at least one subset. We start with an empty pool of subsets (line 1). Then, for each valid requirement $r$ in $R$ (line 2), we attempt to accommodate it in an existing subset in the pool (line 5). If such a subset is found, we include $r$ in the subset (line 6). If not, we populate the pool with an initially empty subset and then include $r$ in the newly added subset (line 12). Note that a subset of requirements $R'$ in this context is specified as the logical conjunction of all the requirements included in the subset, i.e., $\bigwedge_{r' \in R'}r'$. Consequently, to determine whether a new requirement $r$ can be accommodated in an existing subset $R'$, we solve these constraints together with the space model $M$, i.e.,  $r \wedge M \wedge \bigwedge_{r' \in R'}r'$ (line 5), if the resulting constraint is satisfiable then we include $r$ in $R'$ (line 6).

After determining the subsets, for each subset $R'$, we generate a test case by solving the logical conjunction $M \wedge \bigwedge_{r \in R'}r$. The set of test cases generated are then guaranteed to obtain full coverage under the coverage criterion $C$.

\section{Flexibility of U-CIT}
\label{flexibility}

Clearly, this generic algorithm may not be as efficient as their specialized counterparts. For example, it may not generate smaller traditional covering arrays  than the constructors specialized for generating traditional covering arrays. However, it certainly can provide the flexibility needed in practice. In this section we demonstrate the flexibility of U-CIT on a simple, yet (we believe) practical example.

\begin{figure}[t]
 \center
  \includegraphics[width=0.4\textwidth]{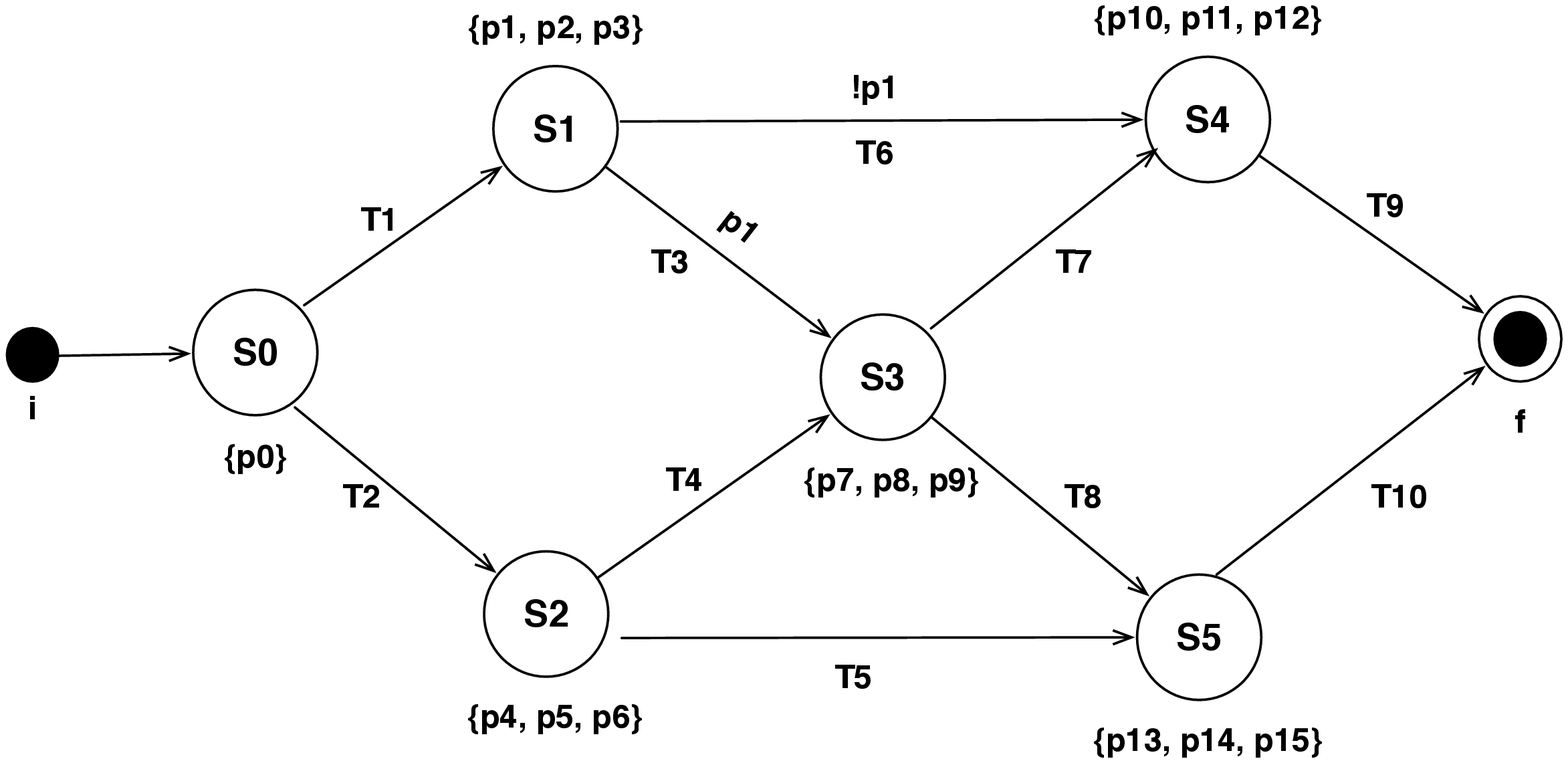}
  \caption{An example finite state machine modeling the behavior of a hypothetical software system.}
  \label{fig:stateDiagram}
\end{figure}

Figure~\ref{fig:stateDiagram} depicts an example finite state machine, modeling the behavior of a hypothetical software system. The model has $6$ states ($S0$-$S5$) together with an initial state $i$ and a final state $f$. Furthermore, we have a total of $16$ boolean factors ($p0$-$p15$). Each factor can be set only in the state in which the factor is defined. Once a factor is set in a state $S$, it is assumed that the factor interacts with any factor defined in a state reachable from $S$. Furthermore, each factor is assumed to be defined in exactly one state. For example, $p1$, $p2$, and $p3$ can be set only in $S1$, but interact with any factors defined in $S3$, $S4$, and $S5$. Some of the transitions are guarded by conditions over factors. For example, transition $T3$ is taken only when $p1$ holds true. Otherwise, transition $T6$ is taken. Moreover, a test case is considered to be a path from the initial to the final state. This machine can, for example, model a mobile application where each state represents a graphical user interface, e.g., a screen, and the factors represent the the boolean input fields defined on screens. 

Consider that developers, using their knowledge of the system, would like to test 1) all possible $3$-way combinations of factor values and 2) all possible $2$-length sequences of the states. They first  attempt to obtain full coverage using traditional covering arrays together with sequence covering arrays. 


To this end, one approach could be to 1) generate a $2$-way sequence covering array for all the states, and 2) for each sequence (from the initial to the final state) selected in the previous step, construct a traditional $3$-way covering array for the factors appearing in the sequence. A sequence covering array that can be computed for this scenario is: \{[$i$,  $S0$, $S2$, $S3$, $S4$, $f$], [$i$, $S0$, $S2$, $S5$, $f$], [$i$, $S0$, $S1$, $S3$, $S5$, $f$], [$i$, $S0$, $S1$, $S3$, $S4$, $f$]\}. Consequently, one issue with this approach is that although the first coverage criterion mandates to cover the $3$-way combinations of values for the factors defined in states $S2$, $S3$, and $S5$ (as there is a path from $S2$ to $S5$), since this path is missing from the sequence covering array, these combinations may never be covered. Thus, this approach does not guarantee to satisfy the first coverage criterion.
  
An alternative approach is to leverage traditional covering arrays first and then attempt to accommodate the required sequences of states. One way is to compute a traditional $3$-way covering array for all the factors. However, this approach will clearly suffer from masking effects~\cite{yilmaz2014reducing}, since there is no path in the machine covering all the states. That is, the factors that cannot interact with each other (e.g., $p1$ and $p4$) will result in invalid test cases, which in turn will prevent some valid combinations from being tested. Another way is to 1) construct a traditional $3$-way covering array for each state, 2) construct a $2$-way sequence covering array for all the states, and 3) for each sequence selected in the previous step, compute the cross product of the traditional covering arrays constructed for the states in the sequence. However, this approach will clearly suffer from severe performance and scalability issues, as there will be many redundant test cases. For example, consider the partial path from $S2$ to $S5$. A $3$-way covering array created for each of these states will have $8$ configurations. Therefore, the cross product of these arrays will have $64$ configurations, the size of which is indeed the same as the exhaustive testing suite of $6$ boolean factors. However, a $3$-way covering array of size $8$ can be created for these $6$ factors. Yet another way is to select the select the paths from the initial to the final state in a ``smart'' manner. For example, the path [$i$, $S0$, $S1$, $S3$, $S4$, $f$] seems to be a good  choice as it is a path of maximum length. Combining all the factors appearing on this path and then generating a $3$-way traditional covering array, on the other hand, will suffer from masking effects due to the overlooked guard condition $p1$ for $T3$; when $!p1$ the interactions between $S0$, $S1$, and $S3$ will not be tested. Unfortunately, this guard condition may not be specified as a system-wide constraint when constructing the traditional covering array, because doing so will invalidate some of the combinations of factor values for $S1$ and $S4$. Note that all combinations of factor values for $S1$ and $S4$ are valid due to $T6$. However, forcing the system always to take $T3$ will prevent some of these combinations from being tested. 

Note that it may be possible to generate specialized approaches for the scenario at hand using traditional and sequence covering array generators.  For example, a feed-back driven adaptive CIT approach, such as the one in~\cite{yilmaz2014reducing}, which keeps on generating valid test cases until all the required combinations and sequences are tested, could be developed.  The point, however, is that it would be a specialized approach, which may or may not be used in other application domains with different types of models and coverage criteria.

One way the proposed approach can handle this scenario is:

\begin{itemize}

\item {\em A U-CIT requirement:} A constraint representing a valid $3$-way combination of factor values or a constraint representing a valid $2$-length sequence of state orderings.

\item {\em A U-CIT test case:} A valid path from the initial state to the final state together with the values of the factors defined in the states appearing on the path.

\item {\em U-CIT space model}: A constraint system specifying 1) the final state machine given in Figure~\ref{fig:stateDiagram} as a system-wide constraint, 2) a valid U-CIT requirement as a U-CIT requirement that does not violate the system-wide constraint, and 3) a U-CIT test case as a U-CIT test case that does not violate the system-wide constraint.

\item {\em U-CIT coverage criterion:} All valid $3$-way combinations of factor values and all valid $2$-length sequences of states must be covered at least once. Determining all valid $2$-length sequences of states can be performed by compiling the given final state machine to a system-wide constraint, and then eliminating all the $2$-length sequences that cannot be satisfied with this system-wide constraint. Determining all valid $3$-tuples additionally requires to determine interacting factors, i.e., factors that can appear on the same path. One way to compute $3$-way combinations of values for interacting factors is to determine all pairs of unreachable states and then remove all valid $t$-tuples involving the factors defined in these states from the valid $t$-way combinations of values for all factors.

\item {\em U-CIT object:} A set of U-CIT test cases that achieves a full coverage under the given U-CIT coverage criterion.

\end{itemize}





\section{Conclusion and Future Work}
\label{conclusion}

We believe that this line of research is novel and can greatly improve the flexibility of combinatorial interaction testing in practice. Therefore, we keep on developing languages and model-based tools for defining variability spaces together with application-specific coverage criteria as well as tools and algorithms for efficiently and effectively computing U-CIT objects.


\end{document}